\documentclass[aps,prd,twocolumn,groupedaddress,amsmath,amssymb]{revtex4-1}
\usepackage{graphicx}  
\usepackage{dcolumn}   
\usepackage{bm}        
\usepackage{verbatim}   
\usepackage{subfigure}
\usepackage[utf8]{inputenc}
\usepackage{color}
\usepackage{braket}
\usepackage{slashed}
\usepackage{amssymb}

\begin{document}

\title{$J/\psi$ photoproduction and polarization in $e+p$ collisions in the improved color evaporation model}
\author{Vincent Cheung}
\affiliation{
   Nuclear and Chemical Sciences Division,
   Lawrence Livermore National Laboratory,
   Livermore, California 94551, USA
   }
\author{Ramona Vogt}
\affiliation{
   Nuclear and Chemical Sciences Division,
   Lawrence Livermore National Laboratory,
   Livermore, California 94551, USA
   }
\affiliation{
   Department of Physics and Astronomy,
   University of California,
   Davis, California 95616, USA
   }
\date{\today}
\begin{abstract}
We calculate the production and polarization of direct $J/\psi$ in the improved color evaporation model in $e+p$ photoproduction. We present the production as functions of transverse momentum, mass of the hadronic final state, and inelasticity. We also present the polarization parameters $\lambda_\vartheta$, $\lambda_{\varphi}$, and $\lambda_{\vartheta \varphi}$ in the helicity and the Collins-Soper frames, as well as the frame-invariant polarization parameter $\tilde{\lambda}$ as a function of transverse momentum and inelasticity. We find agreement with both $J/\psi$ unpolarized cross sections and the invariant polarization parameters as a function of $p_T$.
\end{abstract}


\maketitle


\section{Introduction}

Quarkonium production is important to understand hadronization in QCD. The production mechanism of heavy quarks, from production in hard processes to the hadronization of the final state quarkonium involves both the perturbative and nonperturbative natures of QCD. Nonrelativistic QCD (NRQCD) \cite{Caswell:1985ui} and the color evaporation model \cite{Barger:1979js,Barger:1980mg,Gavai:1994in} remain the most commonly used models today. While NRQCD has difficulties in describing data for $p_T$ cuts less than twice the mass of quarkonium states \cite{Bodwin:2014gia,Faccioli:2014cqa}, both the CEM or the later improved CEM (ICEM) \cite{Ma:2016exq} have only been employed extensively to hadroproduction. In this work, we study the $J/\psi$ production and polarization using the ICEM framework in $e+p$ photoproduction as a timely expansion beyond hadroproduction in preparation for the era of the EIC. It is worth noting that photoproduction in NRQCD has already been calculated at next-to-leading order \cite{Butenschoen:2009zy}, which then was employed to perform a state-of-the-art global fit \cite{Butenschoen:2012qh}. In addition, recent work using potential NRQCD \cite{Brambilla:2022ayc} expands the understanding of the long-distance matrix elements, the model parameters of NRQCD. This work represents the first expansion of (I)CEM beyond hadroproduction, which is needed to test the universality for its model parameter.

Our most recent work extends the ICEM polarization calculation from proton-proton to lead-lead collisions \cite{Cheung:2022nnq}. Despite tensions in $p_T$ dependence of the polarization between our results and the measured ALICE data \cite{ALICE:2011gej,ALICE:2018crw,ALICE:2019lga}, our polarization results generally align with the no polarization experimental conclusion. In this work, we calculate the production and polarization of direct $J/\psi$ production in electron-proton collisions via photoproduction. The effects of feed-down production on $J/\psi$ will be discussed in a later publication.

In this paper, we calculate the polarized direct $J/\psi$ production as a function of transverse momentum ($p_T$), inelasticity ($z$), and mass of the hadronic final state ($W$) in the ICEM using the collinear factorization approach. We present the polarized cross section calculation in Sec.~\ref{polarizedCalculation} and the formation of polarization parameters in Sec.~\ref{polarization-parameters}. The results, along with comparison of unpolarized production distributions to H1 HERA 2 data \cite{H1:2010udv} in Sec.~\ref{unpolarized-production}, and polarization parameters to data in Secs.~\ref{polarization-pt} and ~\ref{polarization-z}. Our conclusions are presented in Sec.~\ref{conclusion}.

\section{Polarized Cross Section}
\label{polarizedCalculation}
The ICEM assumes the $J/\psi$ production cross section takes a constant fraction, $F_{J/\psi}$, of the open $c\bar{c}$ cross section with invariant mass above the mass of the $J/\psi$ but below the hadron threshold, the $D\overline{D}$ pair mass,
\begin{eqnarray}
\sigma_{J/\psi} = F_{J/\psi} \int_{M_{J/\psi}}^{2 m_D} dM \frac{d\sigma}{d M} \biggr\rvert _{p_{c\bar{c}} = \frac{M}{M_{J/\psi}}p_{J/\psi}} \;,
\end{eqnarray}
where a distinction is made between the $c\bar{c}$ momentum and the $J/\psi$ momentum in the ICEM compared to the traditional CEM. In $e+p$ collisions, using the Weizsäcker-Williams approximation \cite{vonWeizsacker:1934nji,Williams:1935dka}, the unpolarized direct $J/\psi$ production cross section in the ICEM is
\begin{eqnarray}
\label{icem-ep-cross-section}
\sigma_{ep\rightarrow J/\psi+X} &=& F_{J/\psi} \sum_{i=q,\bar{q},g} \int^{2m_D}_{M_{J/\psi}}dM dy dQ^2 dx_{i/p} dk_T \nonumber \\
&\times& f_{\gamma/e}(y,Q^2) f_{i/p}(x_i,\mu_F) g(k_T) \nonumber \\
&\times& \sigma_{\gamma + i \rightarrow J/\psi + X} \;,
\end{eqnarray}
where $i$ denotes the parton $q$ ($u, d, s$), $\bar{q}$ ($\bar{u}, \bar{d}, \bar{s}$) or $g$, and $x$ is the momentum fraction of the parton.  Here, $f_{\gamma/e}(y,Q^2)$ is the photon flux for electron-proton collisions of inelasticity, $y$, and virtuality, $-Q^2$, $f_{i/p}(x_i,\mu_F)$ is the parton distribution function for a parton $i$ in the proton as a function of $x_i$ and the factorization scale $\mu_F$.  Finally, $\hat{\sigma}_{\gamma i\rightarrow c\bar{c}+k}$ are the parton-level cross sections for initial states $\gamma + i$ to produce a $c\bar{c}$ pair with a light final-state parton $k$. The invariant mass of the $c\bar{c}$ pair, $M$, is integrated from the physical mass of $J/\psi$ ($M_{J/\psi} =3.10$~GeV) to two times the mass of the $D^0$ hadron ($2m_{D^0} = 3.72$~GeV). Because the $\mathcal{O}(\alpha \alpha_s^2)$ contribution diverges when the light parton is soft, in order to describe the $p_T$ distribution at low $J/\psi$ $p_T$, the initial-state parton is each given a small transverse momentum, $k_T$, kick of $\langle k_T^2\rangle = [1 + (1/12) \ln(\sqrt{s}/20 {\rm ~GeV})] {\rm ~GeV^2}= 1.23{\rm ~GeV}^2$ for $\sqrt{s}=319$~GeV. The collinear parton distribution function is then multiplied by the Gaussian function $g(k_{T})$,
\begin{eqnarray}
g(k_T) &=& \frac{1}{\pi \langle k_T^2 \rangle} \exp (-k_T^2/\langle k_T^2\rangle) \;,
\end{eqnarray}
assuming the $x$ and $k_T$ dependences completely factorize. The same Gaussian smearing is applied in Refs.~\cite{Nelson:2012bc,Ma:2016exq,Mangano:1991jk}. We make the same distinction between the $J/\psi$ and $c\bar{c}$ momenta as adopted in previous ICEM calculations \cite{Ma:2016exq,Cheung:2021epq,Cheung:2022nnq}, which helps describe the $p_T$ distributions at high $p_T$.

We consider production diagrams for photoproduction at $\mathcal{O}(\alpha \alpha_s^2)$ that gives a heavy $c\bar{c}$ pair and a light parton. We denote the momenta of $\gamma$, $i$ , $c$, $\bar{c}$, and $k$ in the partonic process $\gamma+i\rightarrow c+\bar{c} +k$ as $q$, $k_2$, $p_c$, $p_{\bar{c}}$, and $k_3$, respectively, where $k$ is the emitted parton, with $\epsilon_q$, $\epsilon_2$, and $\epsilon_3$ denoting the polarization of the photon and the light partons. When calculating the $2\rightarrow3$ cross section, we transform the momenta of the charm quark ($p_c$) and the anticharm quark ($p_{\bar{c}}$) into the momentum of the proto-$J/\psi$ ($p_{\psi}$) and the relative momentum of the heavy quarks ($k_r$),
\begin{eqnarray}
p_c &=& \frac{1}{2}p_\psi+k_r \;, \\
p_{\bar{c}} &=& \frac{1}{2}p_\psi-k_r \;.
\end{eqnarray}

We note that, since the mass of the proto-$J/\psi$ is integrated from the physical mass of $J/\psi$ to the hadronic threshold, the relative momentum $k_r$ depends on the mass of the proto-$J/\psi$. This differs from NRQCD approaches where the limit $k_r\rightarrow0$ is taken.

We factorize the amplitudes into a product of the color factors, $C$, and the colorless amplitude $\mathcal{A}$, so that $\mathcal{M} = C \mathcal{A}$. The color factors, $C$, in the squared amplitudes are calculated separately by summing over all initial and final state colors, and averaging over the initial state colors.

\begin{widetext}
There are eight diagrams for the $\gamma + g \rightarrow c\bar{c} g$ process.  The factorized amplitudes, $\mathcal{A}$, arranged by the number of three-gluon vertices, are
\begin{eqnarray}
\mathcal{A}_{\gamma g0} &=& i g_\alpha g_s^2 \Bigg[ \bar{u}(p_c)\slashed{\epsilon}_1(\slashed{p}_c-\slashed{q}+m_c)\slashed{\epsilon}_3^*(-\slashed{p}_{\bar{c}}+\slashed{k}_2+m_c)\slashed{\epsilon}_2 v(p_{\bar{c}}) \Bigg] \frac{1}{(p_c - q)^2 - m_c^2 }\frac{1}{(k_2 - p_{\bar{c}})^2 - m_c^2} \nonumber \\
&+& {\rm five~diagrams~with~no~three-gluon~vertices} \;, \\
\mathcal{A}_{\gamma g1} &=& i g_\alpha g_s^2 \Bigg[\bar{u}(p_c)\slashed{\epsilon}_1^*(\slashed{p}_c-\slashed{q}+m_c)\gamma^\mu v(p_{\bar{c}}) \Bigg] [(-2k_2+k_3)\cdot \epsilon_3^* \epsilon_{2\mu} \nonumber \\
&+& (\epsilon_2\cdot\epsilon_3^*) (k_3+k_2)_\mu + (k_2-2k_3)\cdot\epsilon_2 \epsilon_{3\mu}^*)] \frac{1}{(p_c - q)^2 - m_c^2} \frac{1}{(k_2- k_3)^2} \nonumber \\
&+& {\rm one~diagram~with~one~three-gluon~vertex} \;.
\end{eqnarray}

There are four $\gamma+q \rightarrow c\bar{c} q$ diagrams, which are
\begin{eqnarray}
\mathcal{A}_{\gamma q,1} &=& -i g_\alpha g_s^2 \Bigg[ \bar{u}(p_c) \gamma^\nu v(p_{\bar{c}}) \Bigg] \Bigg[ \bar{u}(k_3) \gamma_\nu (\slashed{q} + \slashed{k_2}) \slashed{\epsilon}_q u(k_2) \Bigg] \frac{1}{m_\psi^2}\frac{1}{(q + k_2)^2} \;, \\
\mathcal{A}_{\gamma q,2} &=& -i g_\alpha g_s^2 \Bigg[ \bar{u}(p_c) \slashed{\epsilon}_q (\slashed{p_c} -\slashed{q} + m_c) \gamma^\nu v(p_{\bar{c}}) \Bigg] \Bigg[ \bar{u}(k_3) \gamma_\nu u(k_2) \Bigg] \frac{1}{(p_c-q)^2-m_c^2}\frac{1}{(k_2 - k_3)^2} \;, \\
\mathcal{A}_{\gamma q,3} &=& -i g_\alpha g_s^2 \Bigg[ \bar{u}(p_c) \gamma^\nu (-\slashed{p}_{\bar{c}} + \slashed{q} + m_c) \slashed{\epsilon}_q v(p_{\bar{c}}) \Bigg] \Bigg[ \bar{u}(k_3) \gamma_\nu u(k_2) \Bigg] \frac{1}{(q-p_{\bar{c}})^2-m_c^2}\frac{1}{(k_2 - k_3)^2} \;, \\
\mathcal{A}_{\gamma q,4} &=& -i g_\alpha g_s^2 \Bigg[ \bar{u}(p_c) \gamma^\nu v(p_{\bar{c}}) \Bigg] \Bigg[ \bar{u}(k_3) \slashed{\epsilon}_q  (\slashed{k_3} - \slashed{q}) \gamma_\nu u(k_2) \Bigg] \frac{1}{m_\psi^2}\frac{1}{(q - k_3)^2} \;.
\end{eqnarray}

There are four $\gamma + \bar{q} \rightarrow c\bar{c} \bar{q}$ diagrams, obtained by replacing the spinors in the above five $\gamma + q \rightarrow c\bar{c} q$ diagrams:
\begin{eqnarray}
\mathcal{A}_{\gamma \bar{q},1} &=& -i g_\alpha g_s^2 \Bigg[ \bar{u}(p_c) \gamma^\nu v(p_{\bar{c}}) \Bigg] \Bigg[ \bar{v}(k_3) \gamma_\nu (\slashed{q} + \slashed{k_2}) \slashed{\epsilon}_q v(k_2) \Bigg] \frac{1}{m_\psi^2}\frac{1}{(q + k_2)^2} \;, \\
\mathcal{A}_{\gamma \bar{q},2} &=& -i g_\alpha g_s^2 \Bigg[ \bar{u}(p_c) \slashed{\epsilon}_q (\slashed{p_c} -\slashed{q} + m_c) \gamma^\nu v(p_{\bar{c}}) \Bigg] \Bigg[ \bar{v}(k_3) \gamma_\nu v(k_2) \Bigg] \frac{1}{(p_c-q)^2-m_c^2}\frac{1}{(k_2 - k_3)^2} \;, \\
\mathcal{A}_{\gamma \bar{q},3} &=& -i g_\alpha g_s^2 \Bigg[ \bar{u}(p_c) \gamma^\nu (-\slashed{p}_{\bar{c}} + \slashed{q} + m_c) \slashed{\epsilon}_q v(p_{\bar{c}}) \Bigg] \Bigg[ \bar{v}(k_3) \gamma_\nu v(k_2) \Bigg] \frac{1}{(q-p_{\bar{c}})^2-m_c^2}\frac{1}{(k_2 - k_3)^2} \;, \\
\mathcal{A}_{\gamma \bar{q},4} &=& -i g_\alpha g_s^2 \Bigg[ \bar{u}(p_c) \gamma^\nu v(p_{\bar{c}}) \Bigg] \Bigg[ \bar{v}(k_3) \slashed{\epsilon}_q  (\slashed{k_3} - \slashed{q}) \gamma_\nu v(k_2) \Bigg] \frac{1}{m_\psi^2}\frac{1}{(q - k_3)^2} \;.
\end{eqnarray}
\end{widetext}

We fix the spin of the proto-$J/\psi$ by forming the spin triplet states from the spinor states and extract the orbital angular momentum by projecting the Legendre moments with respect to the relative momentum to obtain the matrix elements
\begin{eqnarray}
    \mathcal{M}_{i_z} = \int d\Omega_k Y_{l=0,m=i_z} \mathcal{M}\;.
\end{eqnarray}

We assume that the angular momentum of the proto-$J/\psi$ is unchanged by the transition from the parton level to the hadron level. We then convolute the partonic cross sections with the CT14 parton distribution functions \cite{Dulat:2015mca} in the domain where $p_\psi \cdot k_r =0$. We restrict the partonic cross section calculations within the perturbative domain by introducing a regularization parameter such that all propagators are at a minimum distance of $Q_{\rm reg}^2=M^2$ from their poles, as employed in Ref.~\cite{Baranov:2002cf} and in the previous ICEM hadroproduction results \cite{Cheung:2021epq}. 

The photoproduction cross section is related to the $e+p$ cross section in Eq.~(\ref{icem-ep-cross-section}) by \cite{Gribov:1962,Budnev:1975poe}
\begin{eqnarray}
\sigma_{\gamma p} &=& \frac{\sigma_{ep}}{\Phi_\gamma}\;,
\end{eqnarray}
where $\Phi_\gamma$ is the effective flux of virtual photons and is computed as
\begin{eqnarray}
\Phi_\gamma &=& \int dy dQ^2 f_{\gamma/e}(y,Q^2) \nonumber \\
&=& \int dy dQ^2 \frac{\alpha}{2\pi y Q^2} \Bigg[ 1+(1-y)^2 - \frac{2M_e^2y^2}{Q^2} \Bigg]\;,
\end{eqnarray}
where $\alpha$ is the electromagnetic coupling constant. To compare to H1 HERA 2 data, the $Q^2$ integral runs from $Q_{min}^2 = M_e^2 y^2/(1-y)$ to 2.5~GeV$^2$, and the $y$ integral runs from $y_{min} = W_{min}^2/s$ to $y_{max} = W_{max}^2/s$. We also take $W_{min} =$~60~GeV, $W_{max} =$~240~GeV, and $\sqrt{s} =$~319~GeV. Contributions from only the transversely polarized photons are considered for the photoproduction calculation. It is worth noting that in electropdocution, where $Q^2$ is much bigger, the longitundinally polarized photons must also be considered as the relative contribution grows as a function of $Q^2$, and thus should have a sizable impact to the polarization of $J/\psi$. To construct the uncertainty bands, we take the factorization and renormalizaton scales to be $\mu_F/m_T = 2.1^{+2.55}_{-0.85}$ and $\mu_R/m_T = 1.6^{+0.11}_{-0.12}$, respectively, where $m_T$ is the transverse mass of the produced charm quark ($m_T = \sqrt{m_c^2+p_T^2}$, where $p_T^2 = 0.5\sqrt{p_{Tc}^2+p_{T\bar{c}}^2}$). We also vary the charm quark mass around $1.27\pm 0.09$~GeV. These variations were determined in Ref.~\cite{Nelson:2012bc}, in which the uncertainties on the total charm cross section were considered.

\section{Polarized production of direct $J/\psi$}
\label{polarization-parameters}
We factor the polarization vector, $\epsilon_\psi(J_z)$, from the unsquared amplitudes for all subprocesses, giving us the form
\begin{eqnarray}
\mathcal{M}_n &=& \epsilon_\psi^\mu(J_z) \mathcal{M}_{n,\mu}
\end{eqnarray}
for each subprocess denoted by the initial states, $n=gg,gq,g\bar{q},q\bar{q}$. The polarization vectors for $J_z=0$, $\pm1$ in the rest frame of the proto-$J/\psi$ are
\begin{eqnarray}
\epsilon_\psi(0)^\mu &=& (1,0,0,0) \;, \\
\epsilon_\psi(\pm1)^\mu &=& \mp\frac{1}{\sqrt{2}}(0,1,\pm i,0) \;,
\end{eqnarray}
using the convention that the fourth component is the $z$ component. While the unpolarized cross section does not depend on the choice of $\hat{R}_z$ axis, the polarized cross sections does depend on the orientation of the $z$ axis. In this calculation, the $\hat{R}_y$ axis is chosen to be the normal vector of the plane formed by the two beams with momenta $\vec{P}_1$ and $\vec{P}_2$,
\begin{eqnarray}
\hat{R}_y &=& \frac{-\vec{P}_1\times\vec{P}_2}{|\vec{P}_1\times\vec{P}_2|} \;.
\end{eqnarray}
In the helicity frame, the $\hat{R}_{z, HX}$ axis is the flight direction of the $c\bar{c}$ pair in the center of mass of the colliding beams. In the Collins-Soper frame \cite{Collins:1977iv}, the $\hat{R}_{z, CS}$ axis is the angle bisector between one beam and the opposite direction of the other beam. The $x$ axis is then determined by the right-handed convention.

We compute the polarized cross section matrix element, $\mathcal{M}_n$, in the rest frame of the $c\bar{c}$ pair by first taking the product of the unsquared amplitude with polarization vector of $J_z=i_z$ corresponding to the polarization axis $\hat{R}_z$ and the unsquared amplitude with polarization vector of $J_z=j_z$ in each subprocess ($n$), then adding them, and finally calculating the components of the polarized cross section matrix according to Eq.~(\ref{icem-ep-cross-section}),
\begin{eqnarray}
\sigma_{i_z,j_z} &=& \int \sum_{n} (\epsilon_\psi^\mu(i_z) \mathcal{M}_{n,\mu})(\epsilon_\psi^{\nu}(j_z) \mathcal{M}_{n,\nu})^* \;,
\end{eqnarray}
where $i_z,j_z=\{-1,0,+1\}$ and the integral is over all variables explicitly shown in Eq.~(\ref{icem-ep-cross-section}) as well as the Lorentz-invariant phase space in $2\rightarrow3$ scatterings. The unpolarized cross section is the trace of the polarized cross section matrix
\begin{eqnarray}
\sigma_{\rm unpol} &=& \sum_{i_z} \sigma_{i_z,i_z} = \sigma_{-1,-1} + \sigma_{0,0} + \sigma_{+1,+1} \;.
\end{eqnarray}

The polarization parameters are calculated using the matrix elements $\sigma_{i_z,j_z}$. The polar anisotropy ($\lambda_{\vartheta}$), the azimuthal anisotropy ($\lambda_\varphi$), and polar-azimuthal correlation ($\lambda_{\vartheta\varphi}$) are given by \cite{Faccioli:2010kd}
\begin{eqnarray}
\lambda_{\vartheta} &=& \frac{\sigma_{+1,+1}-\sigma_{0,0}}{\sigma_{+1,+1}+\sigma_{0,0}} \; \label{lambda_theta_eqn} ,\\
\lambda_{\varphi} &=& \frac{\operatorname{Re}[\sigma_{+1,-1}]}{\sigma_{+1,+1}+\sigma_{0,0}} \;, \\
\lambda_{\vartheta\varphi} &=& \frac{\operatorname{Re}[\sigma_{+1,0}-\sigma_{-1,0}]}{\sqrt{2}(\sigma_{+1,+1}+\sigma_{0,0})} \label{lambda_theta_phi_eqn}\;.
\end{eqnarray}

The polar anisotropy parameter ($\lambda_{\vartheta}$) reflects the proportion of the $J/\psi$ in each spin projection state, with $\lambda_{\vartheta} = 1$ referring to completely transverse production of $J_z=\pm1$ and $\lambda_{\vartheta} = -1$ referring to completely longitudinal production of $J_z=0$.

The azimuthal anisotropy parameter ($\lambda_\varphi$) reflects the azimuthal symmetry of $J/\psi$ production. When $\lambda_\varphi=0$, the production is azimuthally symmetric. When $\lambda_\varphi=\pm1$, the azimuthal distribution is maximally asymmetric. We note that this parameter strongly depends on the production mechanism as well as the measurement frame.

The polar-azimuthal correlation parameter ($\lambda_{\vartheta\varphi}$) describes the angular correlation between $2\vartheta$ and $\varphi$. When $\lambda_{\vartheta\varphi}=0$, the two angles are uncorrelated and as $\lambda_{\vartheta\varphi}$ departs from 0, the behavior of the distribution becomes similar at locations where $2\vartheta=\varphi$. 

The above three polarization parameters depend on the frame (helicity or Collins-Soper) in which they are calculated and measured. However, since the angular distribution itself is rotationally invariant, there are ways to construct invariant polarization parameters from Eqs.~(\ref{lambda_theta_eqn})–(\ref{lambda_theta_phi_eqn}). One of the combinations to form a frame-invariant polarization parameter ($\tilde{\lambda}$) is \cite{Faccioli:2010kd}
\begin{eqnarray}
\tilde{\lambda} &=& \frac{\lambda_\vartheta+3\lambda_\varphi}{1 - \lambda_\varphi} \;.
\end{eqnarray}
The choice of $\tilde{\lambda}$ is the same as the polar anisotropy parameter ($\lambda_\vartheta$) in a frame where the distribution is azimuthally isotropic ($\lambda_{\varphi}=0$). We can remove the frame-induced kinematic dependencies when comparing theoretical predictions to data by also considering the frame-invariant polarization parameter, $\tilde{\lambda}$.

\section{Results}
\label{results-section}
We first show how our approach describes the photoproduction cross section of $J/\psi$ as functions of transverse momentum ($p_T$), inelasticity ($z$), and the mass of the hadronic final state ($W$) by comparing to H1 photoproduction measurements for $Q^2 \lesssim 2.5$~GeV$^2$ at $\sqrt{s}=319$~GeV \cite{H1:2010udv}. We then compare the frame-dependent polarization parameters $\lambda_\vartheta$, $\lambda_{\varphi}$, and $\lambda_{\vartheta\varphi}$ as well as the frame-invariant polarization parameter $\tilde{\lambda}$ to the measured data. In our calculations, we consider theoretical uncertainties by varying the charm quark mass, the renormalization scale, and the factorization scale. The total uncertainty band is constructed by adding the uncertainties in quadrature.

\subsection{Unpolarized $J/\psi$ distributions}
\label{unpolarized-production}

\begin{figure}[t!]
\centering
\includegraphics[width=\columnwidth]{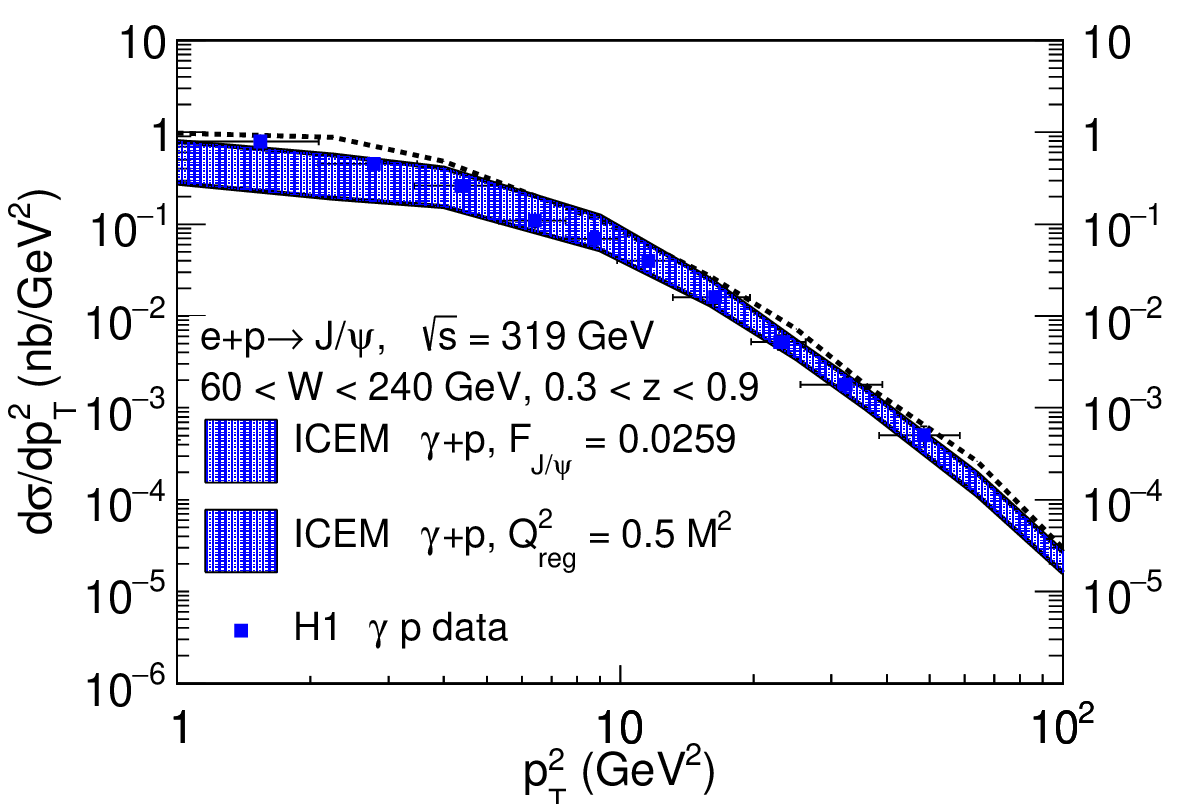}
\caption{The $p_T$ dependence of inclusive $J/\psi$ production at $\sqrt{s} = 319$~GeV with $60<W<240$~GeV and $0.3<z<0.9$ in the ICEM. The combined mass, renormalization scale, and factorization scale uncertainties are shown in the band and compared to the H1 data \cite{H1:2010udv}. The ICEM baseline with $Q_{reg}^2 = 0.5 M^2$ is also shown (dashed). H1 statistical and systematic uncertainties on the data are added in quadrature.} \label{ep-pt-dist}
\end{figure}

We calculate the $p_T$ distribution of direct $J/\psi$ production at $\sqrt{s}=319$~GeV with $60<W<240$~GeV and $0.3<z<0.9$. We assume the direct production is a constant fraction, $0.62$, of the inclusive production \cite{Digal:2001ue} to obtain the inclusive $J/\psi$ $p_T$ distribution. We compare our ICEM inclusive $J/\psi$ $p_T$ distribution with the data measured by the H1 collaboration \cite{H1:2010udv}. The comparison is presented in Fig.~\ref{ep-pt-dist}. By comparing the total cross section for $1<p_T<10$~GeV, we find $F_{J/\psi} = 0.0259$, consistent with previous CEM \cite{Nelson:2012bc} and ICEM calculations in hadroproduction \cite{Cheung:2018tvq, Cheung:2021epq}, where $F_{J/\psi}$ is found to be between 0.0216 and 0.0363.  We add the statistical and systematic uncertainties of the H1 total cross section in quadrature for comparison. We also present the $p_T$ distribution with the regularization parameter, $Q_{reg}^2 = 0.5M^2$ on the same figure to demonstrate effect on the choice of the regularization parameter, which is used to avoid counting the process that would result in a soft gluon being emitted. We will work towards obtaining the next to leading order results in photo production as in done hadroproduction \cite{Petrelli:1997ge}.Overall, we have good agreement with the data over the $p_T$ range measured.

\begin{figure}[t!]
\centering
\includegraphics[width=\columnwidth]{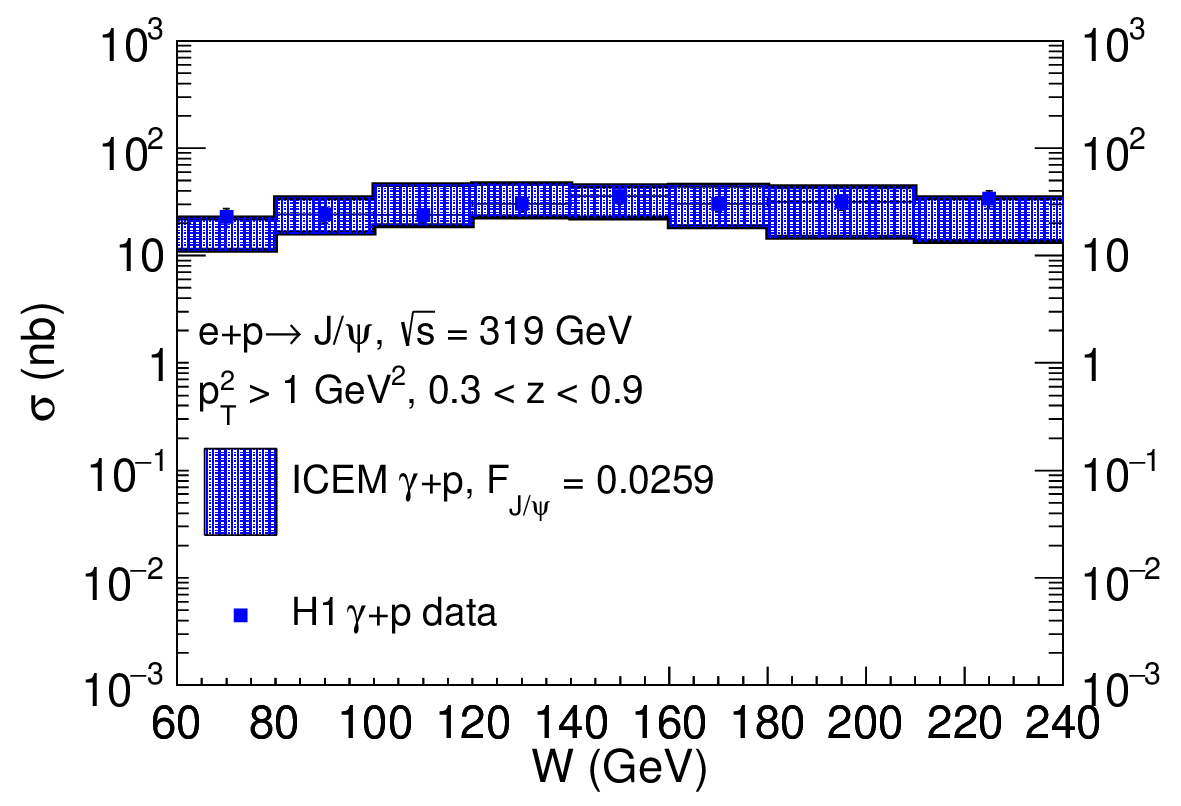}
\caption{The $W$ dependence of inclusive $J/\psi$ production at $\sqrt{s} = 319$~GeV with $p_T^2>1$~GeV and $0.3<z<0.9$ integrated over bins of $W$ in the ICEM. The combined mass, renormalization scale, and factorization scale uncertainties are shown in the band. They are compared to the H1 data \cite{H1:2010udv}. H1 statistical and systematic uncertainties on the data are added in quadrature.} \label{psi_W_data}
\end{figure}

\begin{figure}[b!]
\centering
\includegraphics[width=\columnwidth]{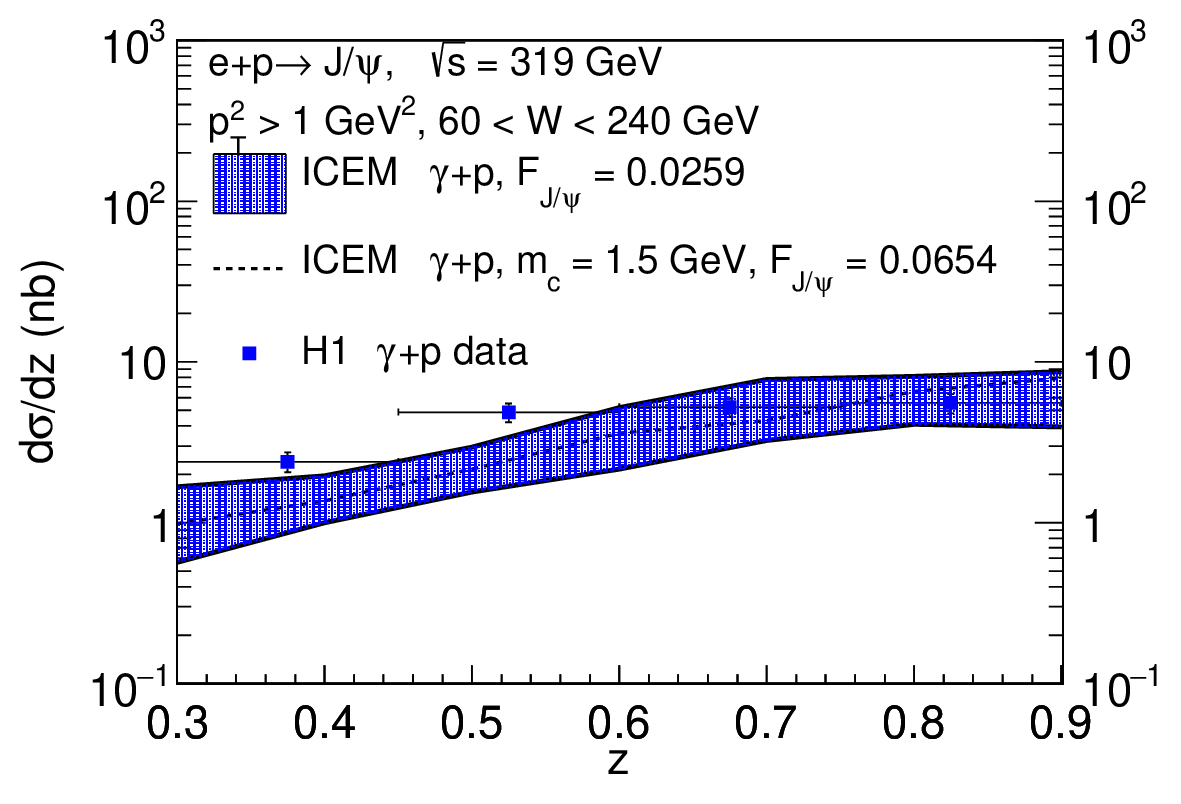}
\caption{The $z$-dependence of inclusive $J/\psi$ production at $\sqrt{s} = 319$~GeV with $1<p_T^2<100$~GeV$^2$ and $60<W<240$~GeV in the ICEM using ($m_c, \mu_F/m_T, \mu_R/m_T$ = 1.27~GeV, 2.1, 1.6) as central result (blue band) and the central result with ($m_c, \mu_F/m_T, \mu_R/m_T$ = 1.5~GeV, 1, 1) in dashed line compared to the H1 data \cite{H1:2010udv}. H1 statistical and systematic uncertainties on the data are added in quadrature.} \label{psi_z_data}
\end{figure}

We present the ICEM inclusive $J/\psi$ photoproduction distributions as a function of the mass of the hadronic final state ($W$) and inelasticity ($z$) in Figs.~\ref{psi_W_data} and \ref{psi_z_data}, respectively. The production in the ICEM in Fig.~\ref{psi_W_data} is integrated over ranges of values of $W$ for comparison. We find the ICEM photoproduction cross section as a function of $W$ to be consistent with the measured result. The production in the ICEM in Fig.~\ref{psi_z_data} slightly underestimates the data in the lowest $z$ bin and slightly overestimates the data at the highest $z$ bin. This trend is also found in photoproduction using NRQCD \cite{Butenschoen:2009zy}. As we have seen in the CEM calculation \cite{Nelson:2012bc} where the uncertainties in total charm and $J/\psi$ cross sections were explored, the best fit values of $\mu_F/m_T$ and $\mu_R/m_T$ depends on the charm mass. We find $F_{J/\psi}$ to be 0.0654 when we choose $m_c$ = 1.5~GeV and $\mu_F/m_T = \mu_R/m_T = 1$. This is also consistent with the findings in Ref. \cite{Nelson:2012bc} where $F_J/psi$ is necessary to be larger for the larger quark mass. Overall, we find the ICEM describes the production data well. We note that all the distributions calculated in the ICEM agree with the NRQCD results \cite{Butenschoen:2009zy}. 

\begin{figure*}
\centering
\begin{minipage}[ht]{0.68\columnwidth}
\centering
\includegraphics[width=\columnwidth]{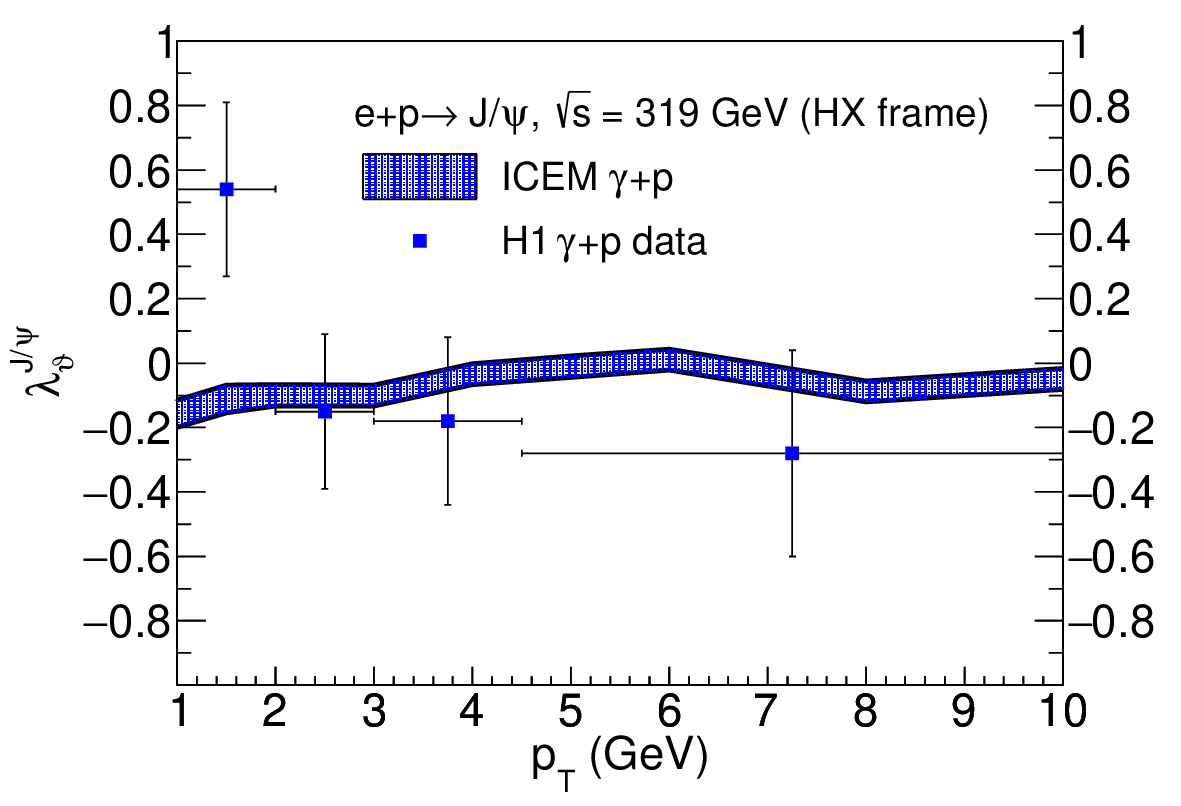}
\end{minipage}%
\begin{minipage}[ht]{0.68\columnwidth}
\centering
\includegraphics[width=\columnwidth]{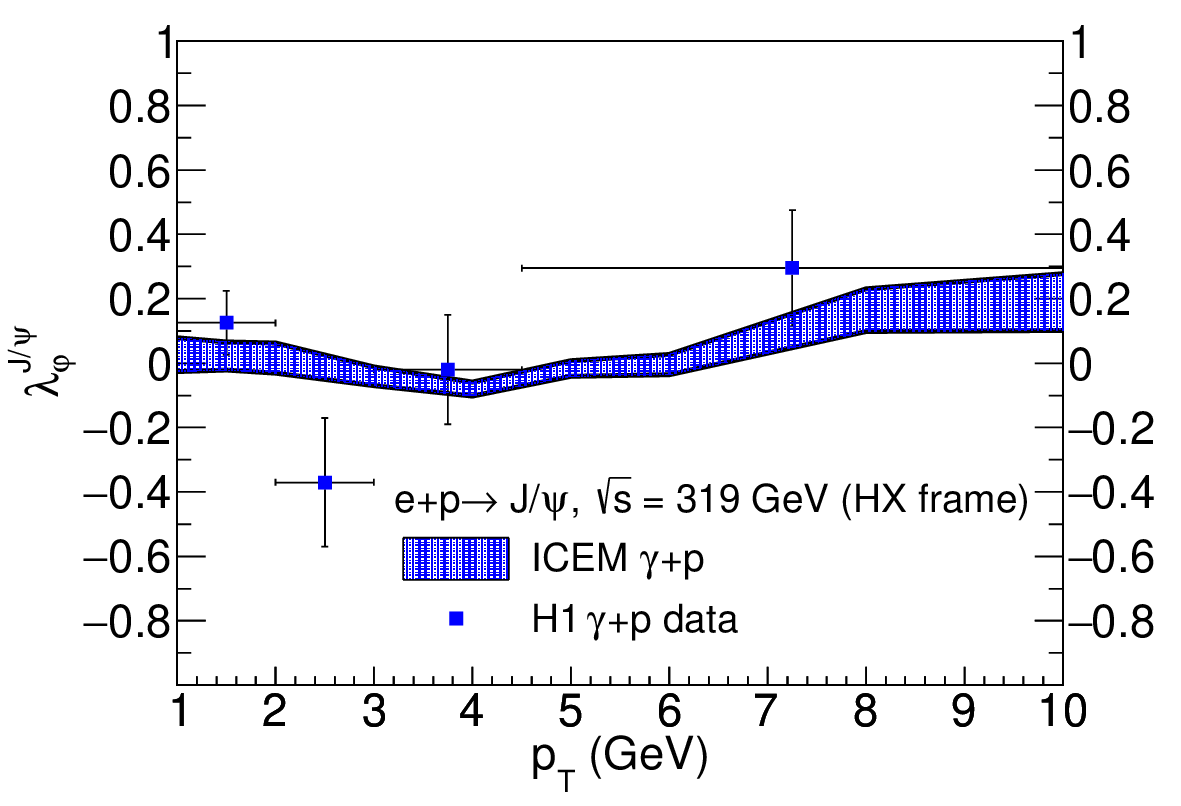}
\end{minipage}
\begin{minipage}[ht]{0.68\columnwidth}
\centering
\includegraphics[width=\columnwidth]{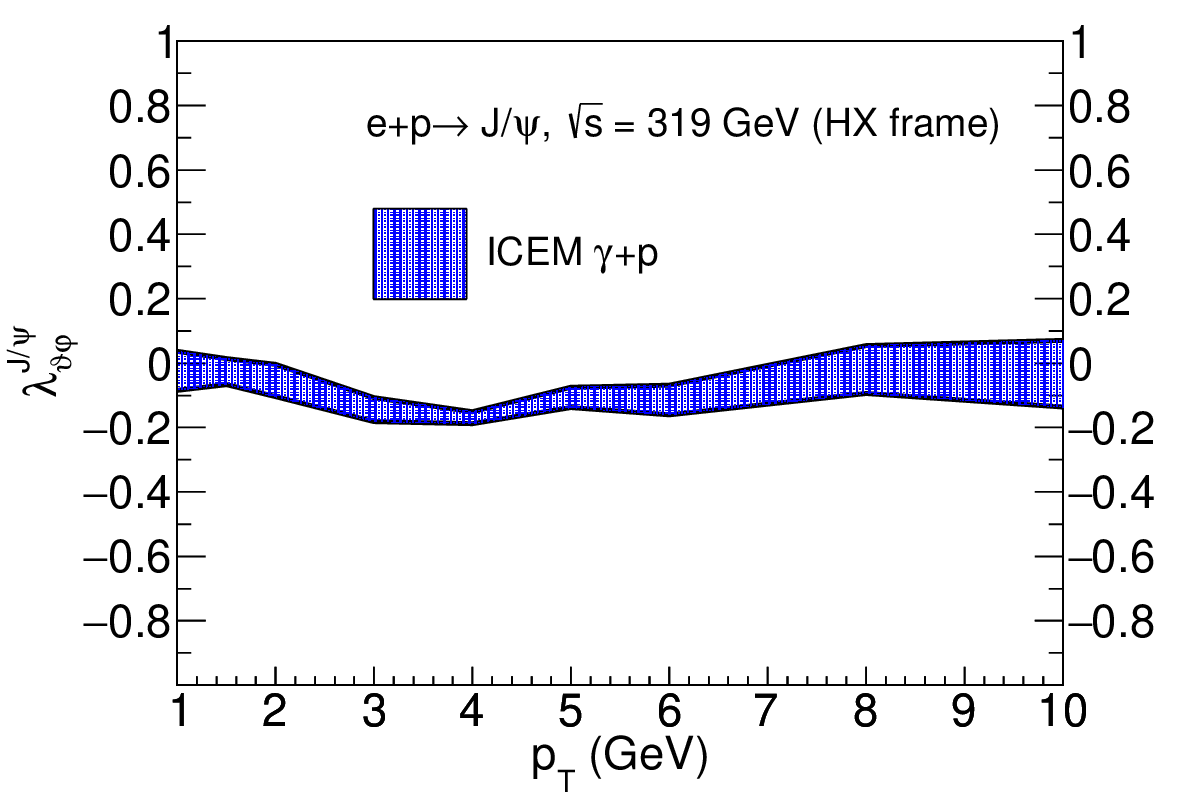}
\end{minipage}
\caption{(a) The polar anisotropy parameter ($\lambda_\vartheta$), (b) the azimuthal anisotropy parameter ($\lambda_\varphi$), and (c) the polar-azimuthal correlation parameter ($\lambda_{\vartheta\varphi}$) as functions of $p_T$ in the helicity frame at $\sqrt{s} = 319$~GeV in the ICEM. The combined mass, renormalization scale, and factorization scale uncertainties are shown in the band and compared to the H1 data \cite{H1:2010udv}.} \label{frame-dependent-lambdas-hx-pt}
\end{figure*}

\begin{figure*}
\centering
\begin{minipage}[ht]{0.68\columnwidth}
\centering
\includegraphics[width=\columnwidth]{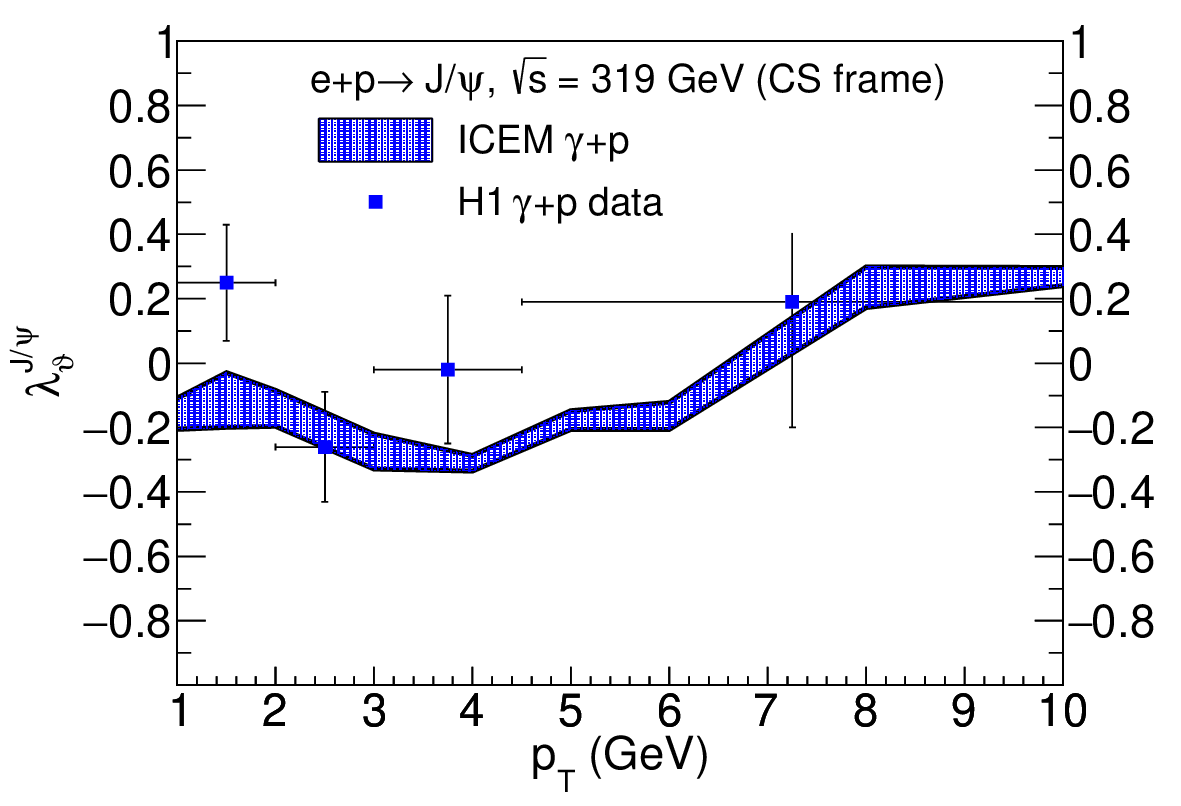}
\end{minipage}%
\begin{minipage}[ht]{0.68\columnwidth}
\centering
\includegraphics[width=\columnwidth]{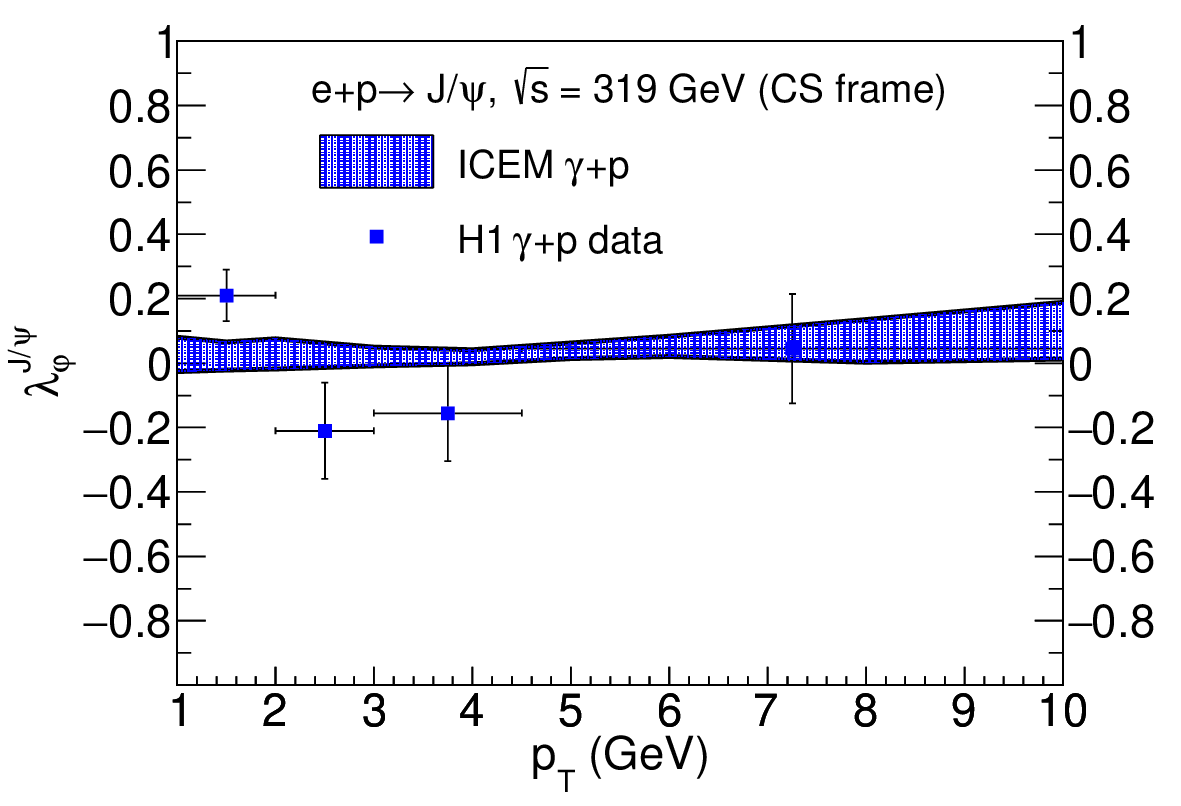}
\end{minipage}
\begin{minipage}[ht]{0.68\columnwidth}
\centering
\includegraphics[width=\columnwidth]{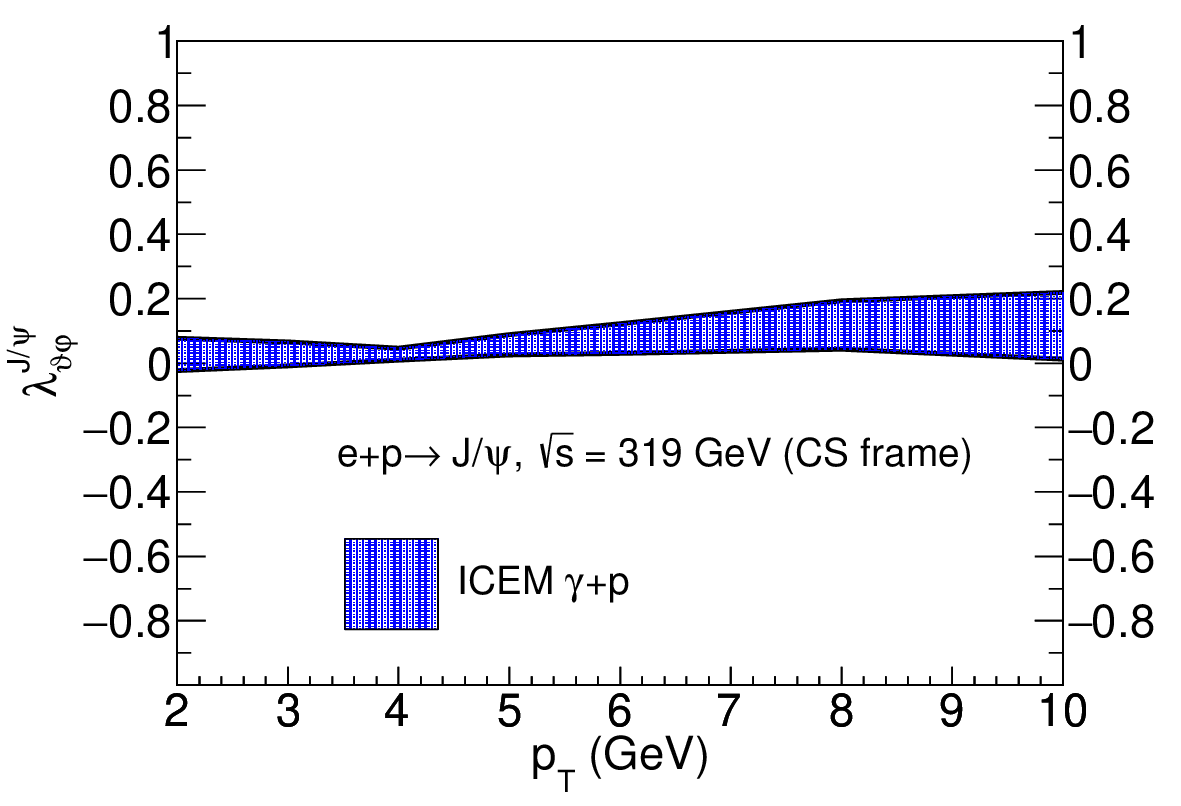}
\end{minipage}
\caption{(a) The polar anisotropy parameter ($\lambda_\vartheta$), (b) the azimuthal anisotropy parameter ($\lambda_\varphi$), and (c) the polar-azimuthal correlation parameter ($\lambda_{\vartheta\varphi}$) as functions of $p_T$ in the Collins-Soper frame at $\sqrt{s} = 319$~GeV in the ICEM. The combined mass, renormalization scale, and factorization scale uncertainties are shown in the band and compared to the H1 data \cite{H1:2010udv}.} \label{frame-dependent-lambdas-cs-pt}
\end{figure*}

\begin{figure}[b!]
\centering
\includegraphics[width=\columnwidth]{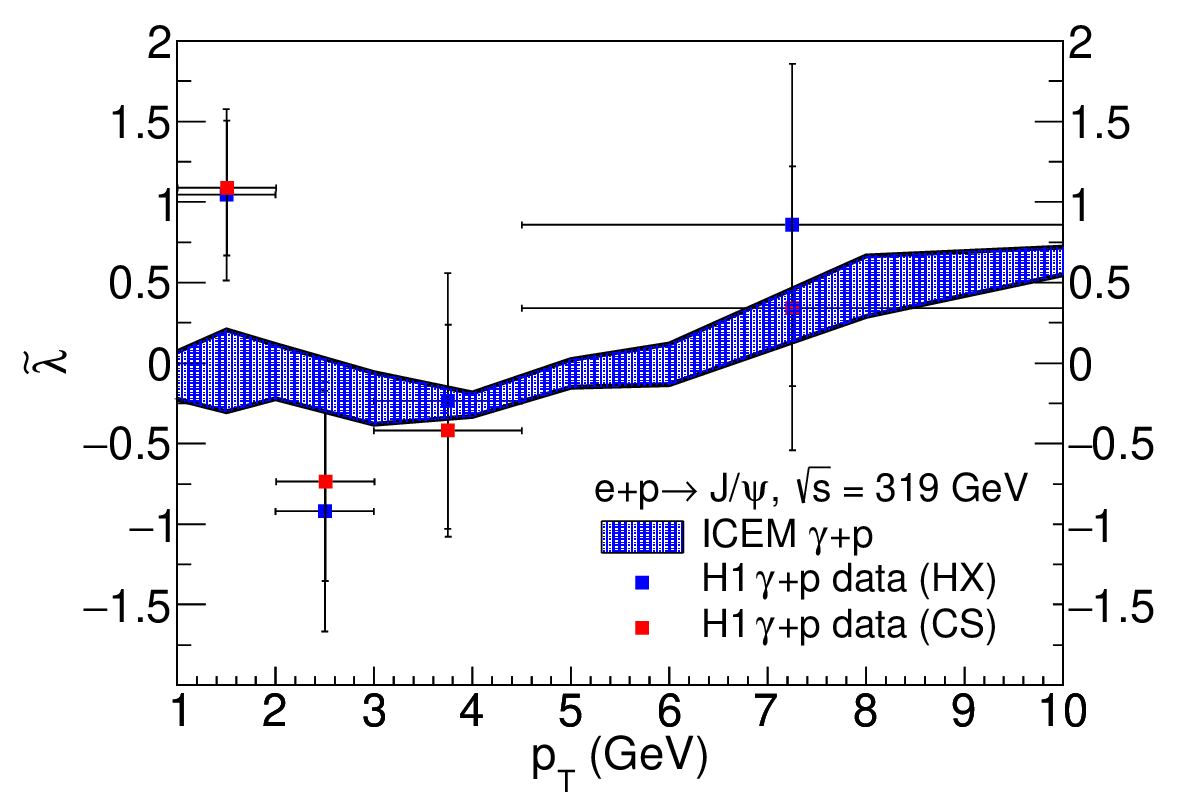}
\caption{The $p_T$-dependence of inclusive $J/\psi$ production at $\sqrt{s} = 319$~GeV as a function of $p_T$ in the ICEM (blue region) compared to the H1 measurements \cite{H1:2010udv} derived from data in the helicity frame (blue squares) and in the Collins-Soper frame (red squares).} 
\label{lambda-invariant-pt}
\end{figure}

\subsection{$p_T$ dependence of $\lambda_\vartheta$, $\lambda_{\varphi}$, $\lambda_{\vartheta\varphi}$, and $\tilde{\lambda}$}
\label{polarization-pt}

We calculate the $p_T$ dependence of the frame-dependent polarization parameters $\lambda_\vartheta$, $\lambda_{\varphi}$, and $\lambda_{\vartheta\varphi}$ at $\sqrt{s}=319$~GeV in the helicity frame and in the Collins-Soper frame. We compare these polarization parameters in the same kinematic ranges presented in Sec. \ref{unpolarized-production} to compare with the data measured by the H1 collaboration \cite{H1:2010udv}. The comparisons in the helicity frame and in the Collins-Soper frame are presented as a function of $p_T$ in Figs. \ref{frame-dependent-lambdas-hx-pt} and \ref{frame-dependent-lambdas-cs-pt}, respectively.

We find $\lambda_{\vartheta}$ is close to zero in both the helicity frame and the Collins-Soper frame at low $p_T$. In the Collins-Soper frame, the transverse component ($J_z = \pm1$) falls off more slowly than the longitudinal component ($J_z = 0)$ as $p_T$ grows. Thus, $\lambda_{\vartheta}$ becomes slightly positive in the Collins-Soper frame but remains near 0 in the helicity frame. 

In both the helicity frame and the Collins-Soper frame, $\lambda_{\varphi}$ is slightly positive, indicating neither the $\hat{R}_{z, CS}$ axis or the $\hat{R}_{z, HX}$ is an symmetry axis of the distribution. $\lambda_{\vartheta\varphi}$ is slightly negative in the helicity frame and is slightly positive in the Collins-Soper frame. We expect discrepancies in the polarization parameters across the two frames as $\hat{R}_{z, CS}$ and $\hat{R}_{z, HX}$ become approximately orthogonal as $p_T$ increases. Overall, the frame-dependent polarization parameters as functions of $p_T$ are generally within the uncertainties of the data.

We show the frame-invariant polarization parameter $\tilde{\lambda}$ as a function of $p_T$ in Fig.~\ref{lambda-invariant-pt}. This removes the frame-dependent kinematics and allows for a quick comparison between our results and the measured data. Only $\lambda_\vartheta$ and $\lambda_\varphi$ are needed to compute the frame-invariant polarization parameter $\tilde{\lambda}$. We also calculate $\tilde{\lambda}$ for the H1 data in the helicity frame and in the Collins-Soper frame.

Since the azimuthal anisotropy parameter $\lambda_\varphi$ is slightly positive, the $p_T$ dependence of the invariant polarization parameter $\tilde{\lambda}$ is positive and becomes significantly transverse at the highest $p_T$. Except for the lowest $p_T$ bin, our photoproduced $J/\psi$-invariant polarization results are in reasonable agreement with the measured data.

\begin{figure*}
\centering
\begin{minipage}[ht]{0.68\columnwidth}
\centering
\includegraphics[width=\columnwidth]{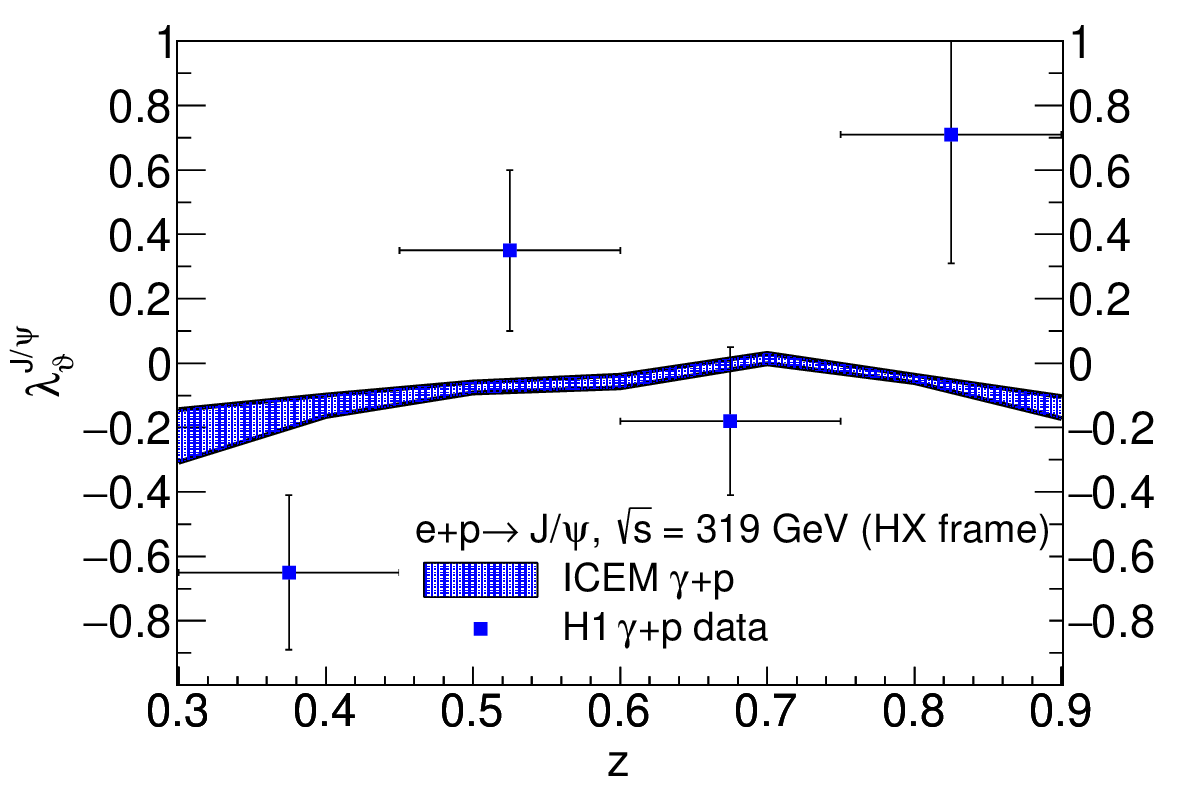}
\end{minipage}%
\begin{minipage}[ht]{0.68\columnwidth}
\centering
\includegraphics[width=\columnwidth]{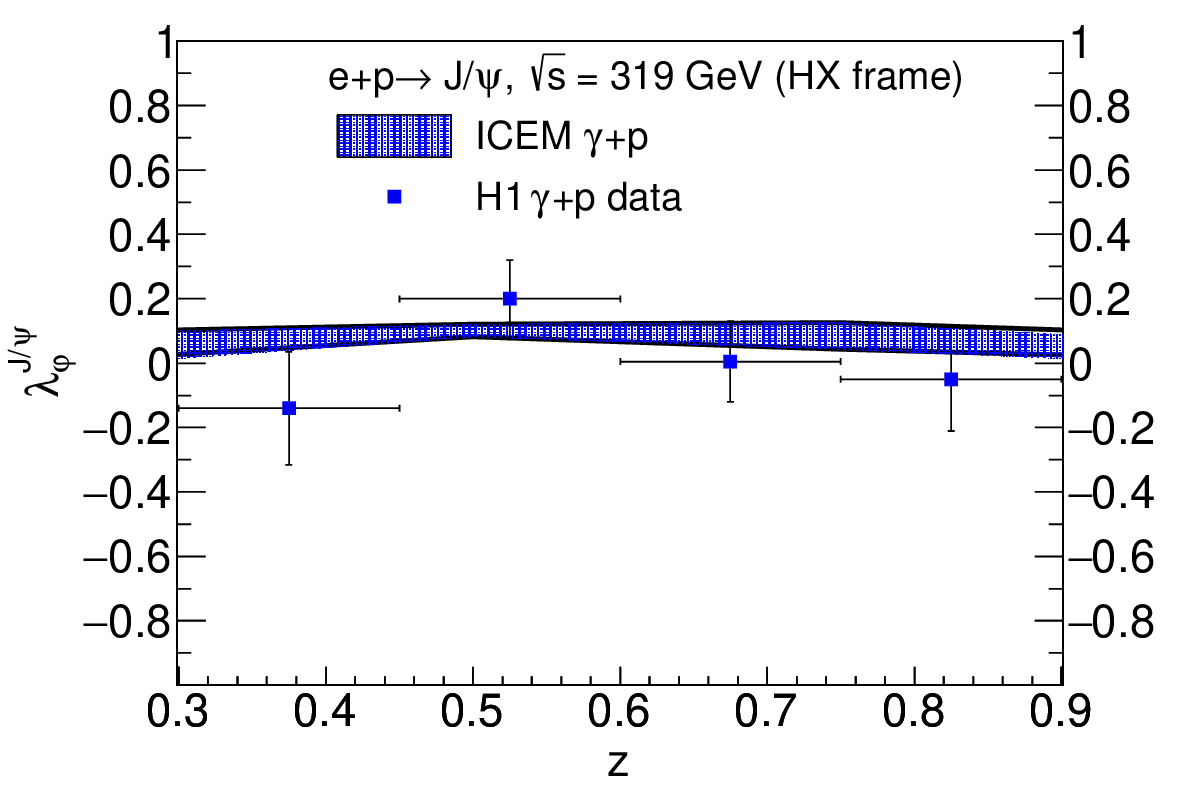}
\end{minipage}
\begin{minipage}[ht]{0.68\columnwidth}
\centering
\includegraphics[width=\columnwidth]{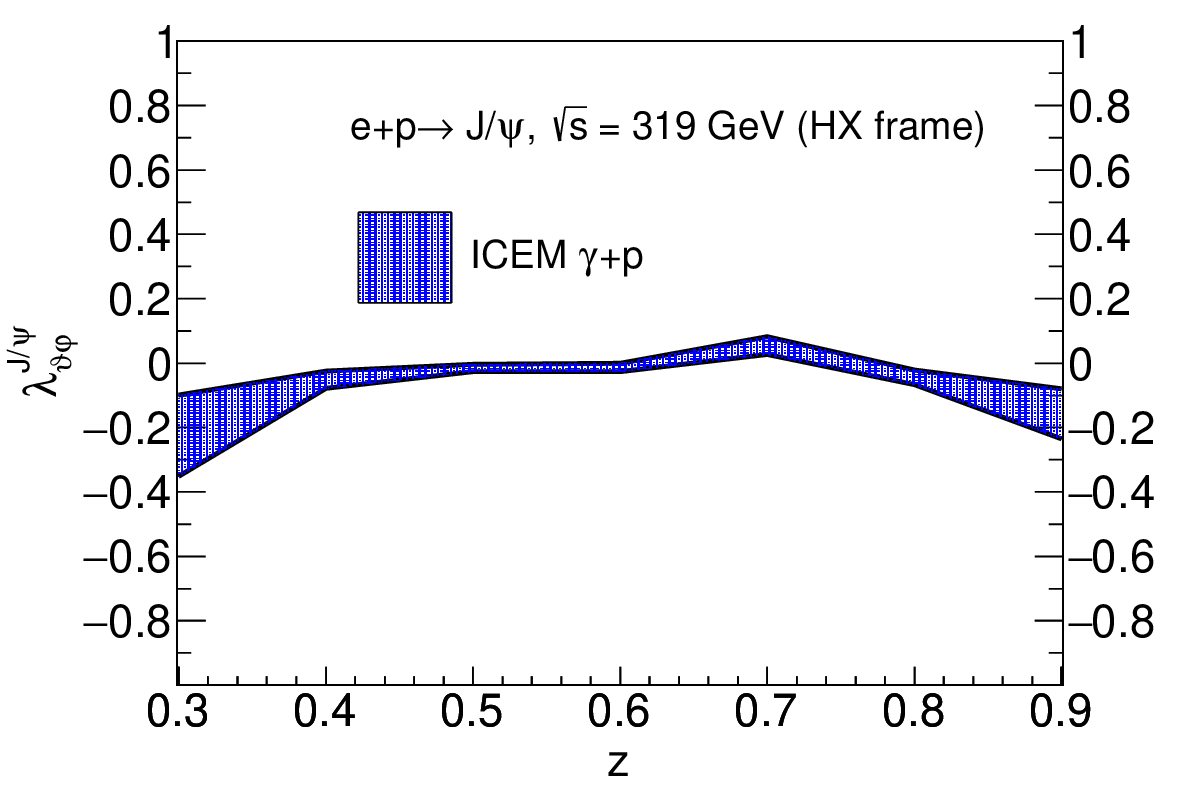}
\end{minipage}
\caption{(a) The polar anisotropy parameter ($\lambda_\vartheta$), (b) the azimuthal anisotropy parameter ($\lambda_\varphi$), and (c) the polar-azimuthal correlation parameter ($\lambda_{\vartheta\varphi}$) as functions of $z$ in the helicity frame at $\sqrt{s} = 319$~GeV in the ICEM. The combined mass, renormalization scale, and factorization scale uncertainties are shown in the band and compared to the H1 data \cite{H1:2010udv}.} \label{frame-dependent-lambdas-hx-z}
\end{figure*}

\begin{figure*}
\centering
\begin{minipage}[ht]{0.68\columnwidth}
\centering
\includegraphics[width=\columnwidth]{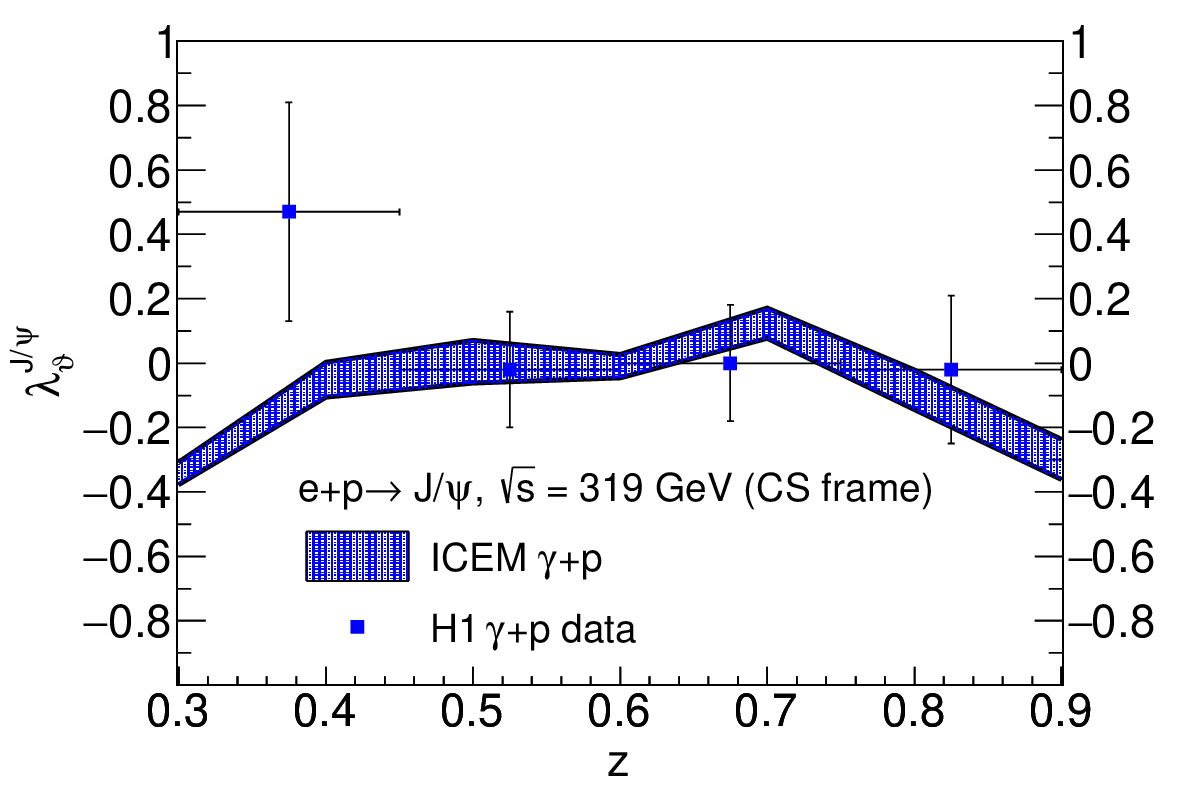}
\end{minipage}%
\begin{minipage}[ht]{0.68\columnwidth}
\centering
\includegraphics[width=\columnwidth]{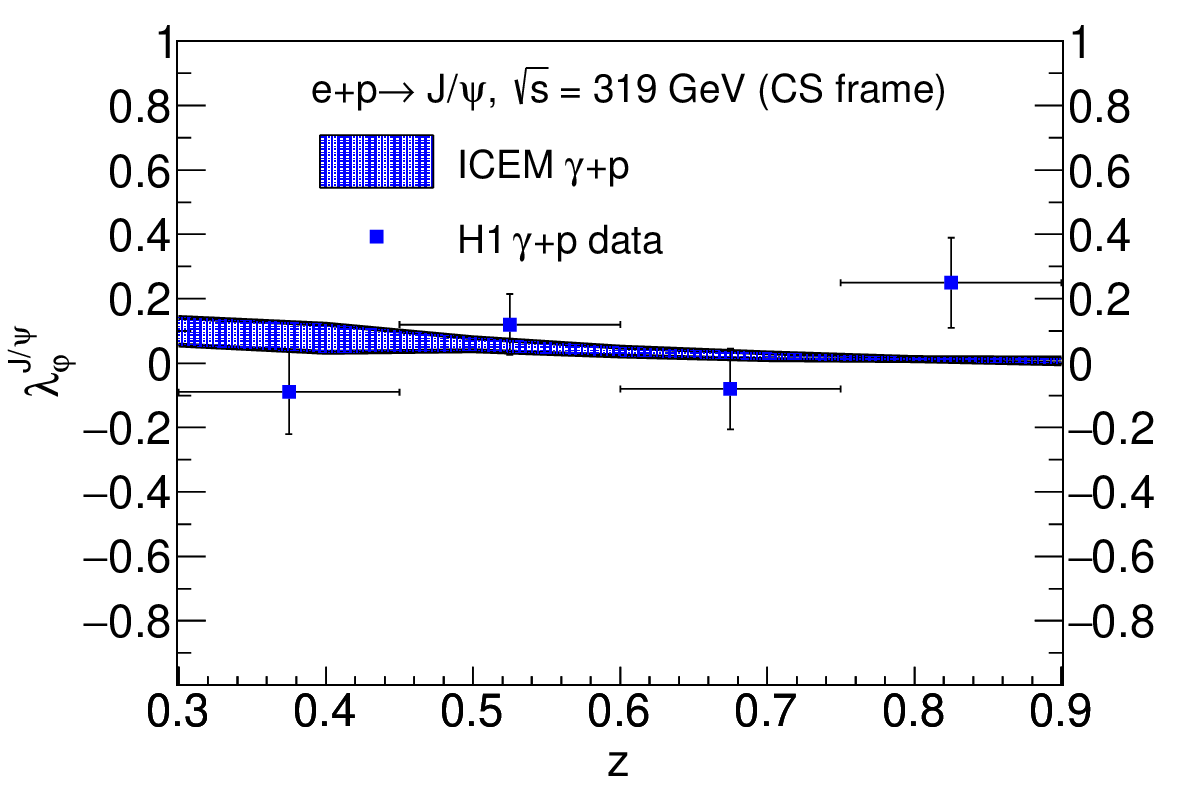}
\end{minipage}
\begin{minipage}[ht]{0.68\columnwidth}
\centering
\includegraphics[width=\columnwidth]{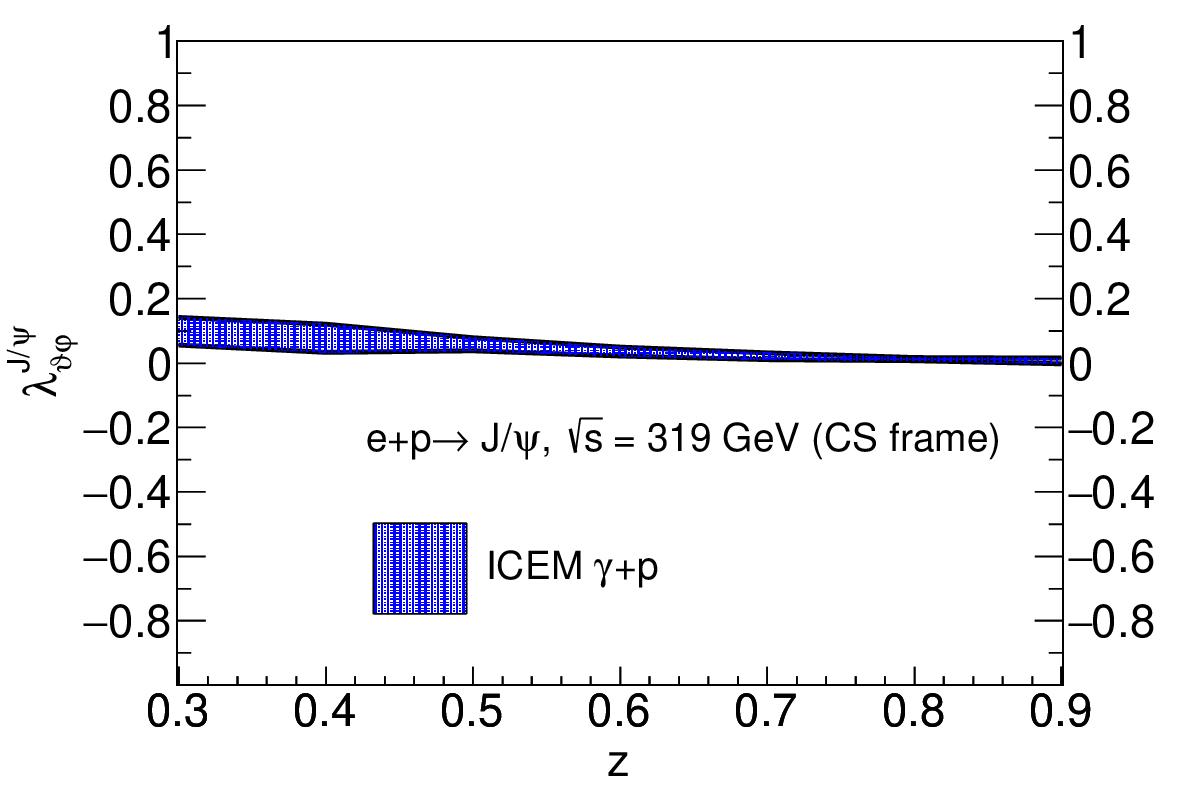}
\end{minipage}
\caption{(a) The polar anisotropy parameter ($\lambda_\vartheta$), (b) the azimuthal anisotropy parameter ($\lambda_\varphi$), and (c) the polar-azimuthal correlation parameter ($\lambda_{\vartheta\varphi}$) (c) as functions of $z$ in the Collins-Soper frame at $\sqrt{s} = 319$~GeV in the ICEM. The combined mass, renormalization scale, and factorization scale uncertainties are shown in the band and compared to the H1 data \cite{H1:2010udv}.} \label{frame-dependent-lambdas-cs-z}
\end{figure*}

\begin{figure}[b!]
\centering
\includegraphics[width=\columnwidth]{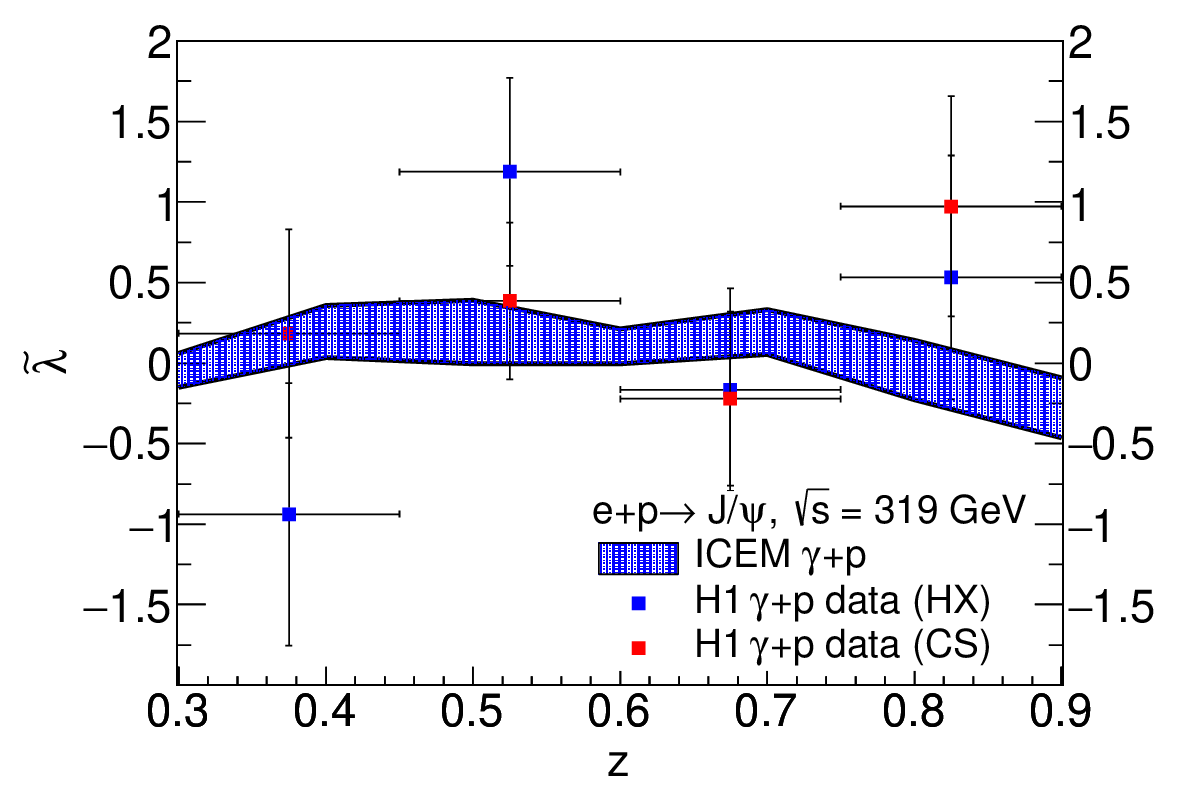}
\caption{The $p_T$ dependence of inclusive $J/\psi$ production at $\sqrt{s} = 319$~GeV as a function of $z$ in the ICEM (blue region) compared to the H1 measurements \cite{H1:2010udv} derived from data in the helicity frame (blue squares) and in the Collins-Soper frame (red squares).} 
\label{lambda-invariant-z}
\end{figure}

\subsection{$z$ dependence of $\lambda_\vartheta$, $\lambda_{\varphi}$, $\lambda_{\vartheta\varphi}$, and $\tilde{\lambda}$}
\label{polarization-z}
We present the $z$ dependence of the frame-dependent polarization parameters in the helicity frame and in the Collins-Soper frame in Figs.~\ref{frame-dependent-lambdas-hx-z} and \ref{frame-dependent-lambdas-cs-z} respectively. We integrate the results from $60<W<240$ and $1<p_T^2<100$~GeV$^2$ to compare with the data measured by the H1 collaboration \cite{H1:2010udv}. 

We find $\lambda_\vartheta$ to be similar in the helicity frame and in the Collins-Soper frame. $\lambda_\vartheta$ is slightly negative in both frames, indicating the production is longitudinal. However, $\lambda_\vartheta$ is more negative in the Collins-Soper frame than in the helicity frame, which is consistent with the dependence at low $p_T$. Agreement with the data is better in the Collins-Soper frame than in the helicity frame.

We find $\lambda_\varphi$ to be slightly positive in both the helicity frame and in the Collins-Soper frame. We find the ICEM results are well within the uncertainties of the data. We do not find $\lambda_{\vartheta\varphi}$ to have a significant $z$ dependence and is consistent with 0. We expect less difference in the polarization parameters across the frames as functions of $z$ because the $p_T$ dependence is integrated over. 

We show the frame-invariant polarization parameter $\tilde{\lambda}$ as a function of $z$ in Fig.~\ref{lambda-invariant-z}. Similarly, we calculate $\tilde{\lambda}$ for the H1 data in the helicity frame and in the Collins-Soper frame for comparison. Even though the ICEM calculation lies within the uncertainties of the data, since the experimental uncertainties on the invariant polarization are large, and the average values of $\tilde{\lambda}$ are substantially different when measured in one frame than the other, it is very hard to draw a conclusion about the invariant polarization as a function of $z$.

\section{Conclusions}
\label{conclusion}
We have presented direct $J/\psi$ photoproduction as functions of $p_T$, $W$, and $z$, as well as the polarization as functions of $p_T$ and $z$ in the improved color evaporation model. We compare our calculations to the unpolarized production and the polarization results measured by the H1 collaboration. We find that the ICEM direct $J/\psi$ photoproduction is consistent with the cross section data with an $F_\mathcal{J/\psi}$ that is also consistent with previous CEM and ICEM calculations. The future Electron-Ion Collider will test the universality of $F_\mathcal{J/\psi}$ at a lower collision energy. We find the invariant polarization to be near unpolarized at small and moderate $p_T$, becoming transverse in the high $p_T$ limit. In the near future we will see if the feed-down contribution will have an influence on the parameter for better agreement with the data. We will study the effects of feed-down production in this approach in a future publication.


\section{Acknowledgments}
This work was performed under the auspices of the U.S. Department of Energy by Lawrence Livermore National Laboratory under Contract No. DE-AC52-07NA27344 and supported by the U.S. Department of Energy, Office of Science, Office of Nuclear Physics (Nuclear Theory) under Contract No. DE-SC0004014 and through the Topical Collaboration in Nuclear Theory on ``Heavy-Flavor Theory (HEFTY) for QCD Matter" under Grant No. DE-SC0023547, and the LLNL-LDRD Program under Project No. 23-LW-036.
\bigskip



\begin{thebibliography}{99}

\bibitem{Caswell:1985ui} 
W.~Caswell and G.~P.~Lepage, Phys.\ Lett.\ {\bf 167B}, 437 (1986).

\bibitem{Barger:1979js} 
V.~D.~Barger, W.~Y.~Keung, and R.~J.~N.~Phillips, Phys.\ Lett.\  {\bf 91B}, 253 (1980).

\bibitem{Barger:1980mg} 
V.~D.~Barger, W.~Y.~Keung, and R.~J.~N.~Phillips, Z.\ Phys.\ C {\bf 6}, 169 (1980).

\bibitem{Gavai:1994in} 
R.~Gavai, D.~Kharzeev, H.~Satz, G.~A.~Schuler, K.~Sridhar, and R.~Vogt, Int.\ J.\ Mod.\ Phys.\ A {\bf 10}, 3043 (1995).

\bibitem{Bodwin:2014gia} 
G.~T.~Bodwin, H.~S.~Chung, U.~R.~Kim, and J.~Lee, Phys.\ Rev.\ Lett.\  {\bf 113}, 022001 (2014).

\bibitem{Faccioli:2014cqa} 
P.~Faccioli, V.~Knünz, C.~Lourenco, J.~Seixas, and H.~K.~Wöhri, Phys.\ Lett.\ B {\bf 736}, 98 (2014).

\bibitem{Ma:2016exq} 
Y.~Q.~Ma and R.~Vogt, Phys.\ Rev.\ D {\bf 94}, 114029 (2016).

\bibitem{Butenschoen:2009zy}
M.~Butenschoen and B.~A.~Kniehl, Phys. Rev. Lett. \textbf{104}, 072001 (2010).

\bibitem{Butenschoen:2012qh}
M.~Butenschoen and B.~A.~Kniehl, Nucl. Phys. B, Proc. Suppl. \textbf{222-224}, 151 (2012).

\bibitem{Brambilla:2022ayc}
N.~Brambilla, H.~S.~Chung, A.~Vairo, and X.~P.~Wang, J. High Energy Phys. 03 (2023) 242.

\bibitem{Cheung:2022nnq}
V.~Cheung and R.~Vogt, Phys. Rev. C \textbf{105}, 055202 (2022).

\bibitem{ALICE:2011gej}
B.~Abelev \textit{et al.} (ALICE Collaboration), Phys. Rev. Lett. \textbf{108}, 082001 (2012).

\bibitem{ALICE:2018crw}
S.~Acharya \textit{et al.} (ALICE Collaboration), Eur. Phys. J. C \textbf{78}, 562 (2018).

\bibitem{ALICE:2019lga}
S.~Acharya \textit{et al.} (ALICE Collaboration), J. High Energy Phys. 02 (2020) 041.

\bibitem{H1:2010udv}
F.~D.~Aaron \textit{et al.} (H1 Collaboration), Eur. Phys. J. C \textbf{68}, 401 (2010).

\bibitem{vonWeizsacker:1934nji}
C.~F.~von Weizsacker, Z. Phys. \textbf{88}, 612 (1934).

\bibitem{Williams:1935dka}
E.~J.~Williams, Kong. Dan. Vid. Sel. Mat. Fys. Med. \textbf{13N4}, 1 (1935).

\bibitem{Nelson:2012bc} 
R.~E.~Nelson, R.~Vogt, and A.~D.~Frawley, Phys.\ Rev.\ C {\bf 87}, 014908 (2013).

\bibitem{Mangano:1991jk}
M.~L.~Mangano, P.~Nason and G.~Ridolfi, Nucl. Phys. \textbf{B373}, 295 (1992).

\bibitem{Cheung:2021epq}
V.~Cheung and R.~Vogt, Phys. Rev. D \textbf{104}, 094026 (2021).

\bibitem{Petrelli:1997ge}
A.~Petrelli, M.~Cacciari, M.~Greco, F.~Maltoni and M.~L.~Mangano, Nucl. Phys. \textbf{B514}, 245 (1998).

\bibitem{Dulat:2015mca}
S.~Dulat, T.~J.~Hou, J.~Gao, M.~Guzzi, J.~Huston, P.~Nadolsky, J.~Pumplin, C.~Schmidt, D.~Stump, and C.~P.~Yuan, Phys. Rev. D \textbf{93}, 033006 (2016).

\bibitem{Baranov:2002cf} 
S.~P.~Baranov, Phys.\ Rev.\ D {\bf 66}, 114003 (2002).

\bibitem{Gribov:1962}
V.~N.~Gribov \textit{et al.}, Sov. Phys. JETP \textbf{14}, 1308 (1962).

\bibitem{Budnev:1975poe}
V.~M.~Budnev, I.~F.~Ginzburg, G.~V.~Meledin, and V.~G.~Serbo, Phys. Rep. \textbf{15}, 181 (1975).

\bibitem{Collins:1977iv} 
J.~C.~Collins and D.~E.~Soper, Phys.\ Rev.\ D {\bf 16}, 2219 (1977).

\bibitem{Faccioli:2010kd} 
P.~Faccioli, C.~Lourenco, J.~Seixas, and H.~K.~Wohri, Eur.\ Phys.\ J.\ C {\bf 69}, 657 (2010).

\bibitem{Digal:2001ue} 
S.~Digal, P.~Petreczky, and H.~Satz, Phys.\ Rev.\ D {\bf 64}, 094015 (2001).

\bibitem{Cheung:2018tvq} 
V.~Cheung and R.~Vogt, Phys.\ Rev.\ D {\bf 98}, 114029 (2018).

\end{thebibliography}
\end{document}